\documentclass[letterpaper,twocolumn,nonatbib,10pt]{article}
\usepackage{usenix,graphicx,cite,url}
\begin{document}
\fontsize{10}{12.03}
\selectfont

\title{Security Through Amnesia: A Software-Based Solution to the
  Cold Boot Attack on Disk Encryption}

\author{
{\rm Patrick Simmons}\\
University of Illinois at Urbana-Champaign
} 

\maketitle


\abstract Disk encryption has become an important security measure for
a multitude of clients, including governments, corporations,
activists, security-conscious professionals, and privacy-conscious
individuals.  Unfortunately, recent research has discovered an
effective side channel attack against any disk mounted by a running
machine\cite{princetonattack}.  This attack, known as the cold boot
attack, is effective against any mounted volume using state-of-the-art
disk encryption, is relatively simple to perform for an attacker with
even rudimentary technical knowledge and training, and is applicable
to exactly the scenario against which disk encryption is primarily
supposed to defend: an adversary with physical access.  To our
knowledge, no effective software-based countermeasure to this attack
supporting multiple encryption keys has yet been articulated in the
literature.  Moreover, since no proposed solution has been implemented
in publicly available software, all general-purpose machines using
disk encryption remain vulnerable.  We present Loop-Amnesia, a
kernel-based disk encryption mechanism implementing a novel technique
to eliminate vulnerability to the cold boot attack.  We offer
theoretical justification of Loop-Amnesia's invulnerability to the
attack, verify that our implementation is not vulnerable in practice,
and present measurements showing our impact on I/O accesses to the
encrypted disk is limited to a slowdown of approximately 2x.
Loop-Amnesia is written for x86-64, but our technique is applicable to
other register-based architectures.  We base our work on loop-AES, a
state-of-the-art open source disk encryption package for Linux.
\endabstract

\section{Introduction}

The theft of sensitive data from computers owned by governments,
corporations, and legal and medical professionals has escalated to a
problem of paramount importance as computers are now used to store,
modify, and safeguard all kinds of sensitive and private information.
Hard drive thefts in the past have put information such as medical
data\cite{dumbdoc}, Social Security and passport
numbers\cite{wackenhut}, and the access codes for a financial service
corporation's private Intranet\cite{bbcscary}\footnote{In this case,
  the hard drive appears to have been sold and purchased legitimately
  but was not adequately wiped prior to the sale.} at risk.

Because of the significant potential for harm such breaches represent,
disk encryption, in which an entire filesystem is stored on
nonvolatile storage in encrypted form, has become a standard and often
mandatory security technique in many environments\cite{bureaucracy}.
Most major commercial operating systems now offer some form of
kernel-based disk encryption\cite{linux}\cite{solaris}\cite{windows},
and third-party tools supporting disk encryption, such as
TrueCrypt\cite{truecrypt}, are freely available for many architectures
and operating systems.  This software has proven effective against
determined adversaries wishing to defeat its protection\cite{boucher}.

However, recent work\cite{princetonattack} by Halderman et al.\ has
uncovered a flaw common to all commercially available disk encryption
packages.  These researchers observed that, as long as an encrypted
volume is mounted, a disk encryption package will store the encryption
key in RAM.  They further discovered that, contrary to popular belief,
DRAM does not lose its contents for several minutes after a loss of
power.  Thus, Halderman et al.\ put forth the following attack on all
disk encryption: cut power to the target machine, pull out the RAM,
put the RAM in a new machine\footnote{Variants of the attack eliminate
  the need for a separate machine.}, and boot this machine with an
attack program of their creation which overwrites a minimal amount of
RAM with its own code and dumps the original contents of RAM to
nonvolatile storage.  At this point, an attacker can search the
contents of RAM for the encryption key or simply try every key-length
string of bits present in the RAM of the original machine as a
potential key.  This attack, called the ``cold-boot attack'' by
Halderman et al., is simple to perform, routinely effective, and
broadly applicable against the existing universe of disk encryption
software packages.

It is difficult to overstate the significance of the cold-boot attack.
The protection afforded by disk encryption against any adversary with
access to the running target machine is now effectively skewered.
Many users for whom disk encryption previously offered protection are
now at risk of having their data stolen when their machines are stolen
or lost.  One may argue that these users should physically secure
their machines, but, as disk encryption is specifically intended to
protect against an attacker who has physical access to the disk, that
argument rings hollow.

In this paper, we describe the novel implementation approach we used
in Loop-Amnesia, the first disk encryption software package not
vulnerable to the cold-boot attack.  We contribute a method of
permanently storing an encryption key inside CPU registers rather than
in RAM, an approach of capitalizing on this ability to allow the
masking of arbitrarily many encryption keys from disclosure under a
cold-boot attack, an implementation strategy for the AES encryption
algorithm which ensures that no data related to encryption keys is
ever leaked to RAM, a prototype implementation of our approach, and
performance measurements validating our technique's usability in
practice.

Section 2 describes the attack model used by our paper.  Section 3
provides an overview of AES and the loop-AES software package we
enhanced to thwart the cold-boot attack.  Section 4 describes the
design of Loop-Amnesia.  Section 5 describes our implementation.
Section 6 describes our justification that Loop-Amnesia is in fact
immune to the cold-boot attack and describes our correctness testing.
Section 7 details our performance benchmarking of Loop-Amnesia.
Section 8 details the limitations of our approach to this problem.
Section 9 describes related work.  Section 10 describes future work.
Section 11 concludes the paper.

\section{Attack Model}

We assume our attacker has full physical access to the target machine.
The attacker is assumed to possess any commonly available equipment
necessary or useful for performing the cold-boot attack, such as his
own computer or other device capable of reading RAM after he has
removed it from the target machine.

In the event our attacker has access to an account on the target
machine, such as with stolen login credentials or due to the fact that
the machine was stolen with a user logged in, we seek to prevent the
attacker from gaining unauthorized access to the disk volume key or to
parts of the encrypted disk to which the account he is using does not
have access.  We assume an attacker will not be able to gain access to
the encryption keys through vulnerabilities in the operating system;
other work (SVA\cite{sva}, SECVisor\cite{secvisor}, and
HyperSafe\cite{hypersafe}) can protect the kernel from exploitation.

\section{Background}

\subsection{Aspects of AES Relevant to Loop-Amnesia}

AES, or the Advanced Encryption Standard, is an efficient block cipher
algorithm.  Originally published as Rijndael\cite{rijndael}, the
algorithm became the AES standard in 2001.  It has proven quite
resistant to
cryptanalysis\cite{cryptoeprint:2009:317}\cite{cryptoeprint:2009:531}
since its standardization.

\subsubsection{Rounds}

AES encryption proceeds in multiple \emph{rounds}.  In a round-based
encryption process, plaintext is first encoded to ciphertext by
applying the main body of the encryption algorithm.  The resulting
ciphertext is then encrypted again using the same algorithm in a
second round of encryption.  This process is repeated a number of
times: in the case of 128-bit AES, our algorithm of concern, the
number of rounds is 10.

\subsubsection{Key Schedule}

In order to increase the algorithm's resistance to cryptanalysis, AES
and other block ciphers employ a concept called a \emph{key schedule},
in which a different key is used for each round of encryption.  In
AES, the original key is used for the first round, and subsequent
rounds use keys obtained by permuting the contents of the previous
round key.  This permutation is reversible.  In most AES
implementations, all 10 keys of the key schedule are precomputed and
stored to RAM for performance purposes.\footnote{For reasons discussed
  in later sections, this performance optimization is foreclosed to
  Loop-Amnesia.}  Since there are different but related key schedules
for encryption and decryption, a total of 20 128-bit quantities from
which the original key can be derived are stored to RAM when using
unmodified loop-AES or a similar disk encryption package.

\subsection{Organization of loop-AES}

Loop-AES\cite{loopaes} is a kernel plugin for Linux providing an
\emph{encrypted loopback device} to the user.  An encrypted loopback
device binds to a normal block device, such as a disk partition or
file, and provides a view of its data after having been decrypted with
a key.  If data is written to the loopback device, it is
encrypted before being stored on the device to which the loopback
device is bound.

The internal structure of loop-AES is both clean and modular.  All
encryption, decryption, and key-setting work is performed by the three
methods \texttt{aes\_encrypt}, \texttt{aes\_decrypt}, and
\texttt{aes\_set\_key}.  Key data is stored inside the
\texttt{aes\_context} structure, which is treated as opaque by all of
loop-AES outside of the three routines mentioned above.  IV
computation, CBC chaining, and other functions necessary to a full
disk encryption system are handled independently of the implementation
of these functions and, indeed, independently of the cryptographic
algorithm used.  Loop-Amnesia's changes to loop-AES are confined to
these three subroutines.

Of particular concern to us is how loop-AES stores cryptographic keys.
Keys are stored only inside the aforementioned opaque
\texttt{aes\_context} structures; loop-AES conscientiously deletes
them from other locations in memory after initializing the
\texttt{aes\_context} structures with \texttt{aes\_set\_key}.  Because
the keys are stored in memory by \texttt{aes\_set\_key}, however,
loop-AES, like other prior disk encryption software, is fully
vulnerable to the cold-boot attack.

\section{The Design of Loop-Amnesia}

The basic insight of Loop-Amnesia's design is that, because of the
ubiquity of model-specific registers, or MSRs, in CPU architectures
today, it is possible to store data inside the CPU, rather than in
RAM, thus making that data unreadable to a perpetrator of the
cold-boot attack.  The challenging aspect of this approach is finding
model-specific registers that can practicably be used for this task:
if an MSR is repurposed as storage space for an encryption key, it is
unavailable for its intended use.  Model-specific registers are used
for a diverse variety of system tasks; some, like the control for the
CPU fan, must not be tampered with lightly lest the safe operation of
the hardware be threatened.

On our target platform, x86-64, we disabled performance counting and
therefore were able to use the performance counter registers to hold a
single 128-bit AES key.\footnote{On Intel\cite{inteldoc} processors,
  we use MSRs 0xC1, 0xC2, 0x309, and 0x30A.  On AMD\cite{amddoc} CPUs,
  we use MSRs 0xC0010004, 0xC0010005, 0xC0010006, and 0xC0010007.}  To
evaluate the generality of our approach, we examined the CPU system
programming manual for a PowerPC chip\cite{freescaledoc}.  We were
also able to find performance counter MSRs on PowerPC that would
appear to be repurposable for key storage on that
architecture.\footnote{However, the manual also states that the
  performance counters are readable from user mode, and it does not
  appear that the instruction to read them can be disabled by the
  operating system.  Thus, our approach may not provide security
  against an attacker with the ability to execute arbitrary user-level
  code on PowerPC unless we found other repurposable MSRs.  On x86-64,
  the ability of unprivileged code to read performance counters is
  configurable by the operating system, and we disable this ability.}
Of course, on any platform, disabling and repurposing the hardware
performance counter infrastructure in this manner has the side effect
of foreclosing the use of any hardware-assisted performance profilers.
Since we expect protection against cold-boot attacks to be most
important for production machines, which do not typically use
hardware-assisted performance profilers, we do not consider this a
serious deficiency of our approach.

Since storing the disk volume key in the MSRs directly would prevent
the mounting of more than one encrypted volume
simultaneously\footnote{Another motivation for supporting multiple
  simultaneous encryption keys is to support a mode of loop-AES which
  uses 64 different encryption keys to protect against watermark
  attacks\cite{loopaes}}, we instead store a randomly generated number
in the MSRs, then use this master key to encrypt the disk volume key
for each mounted volume.  Because we assume an attacker may later have
access to all RAM, we require a random number generator (RNG) which
guarantees that previously output random numbers cannot be calculated
from its subsequent internal state.\footnote{In our implementation, we
  use the Linux kernel random number generator, which is specifically
  designed to provide this guarantee.  There has been some
  cryptanalysis of the Linux RNG with respect to its ability to
  provide this guarantee\cite{linuxrngvuln}, but the implementation is
  still considered safe in practice\cite{linuxrngcounter}.}

\section{Implementation}
\label{implicate}

\begin{figure*}[t]
\includegraphics[scale=0.60]{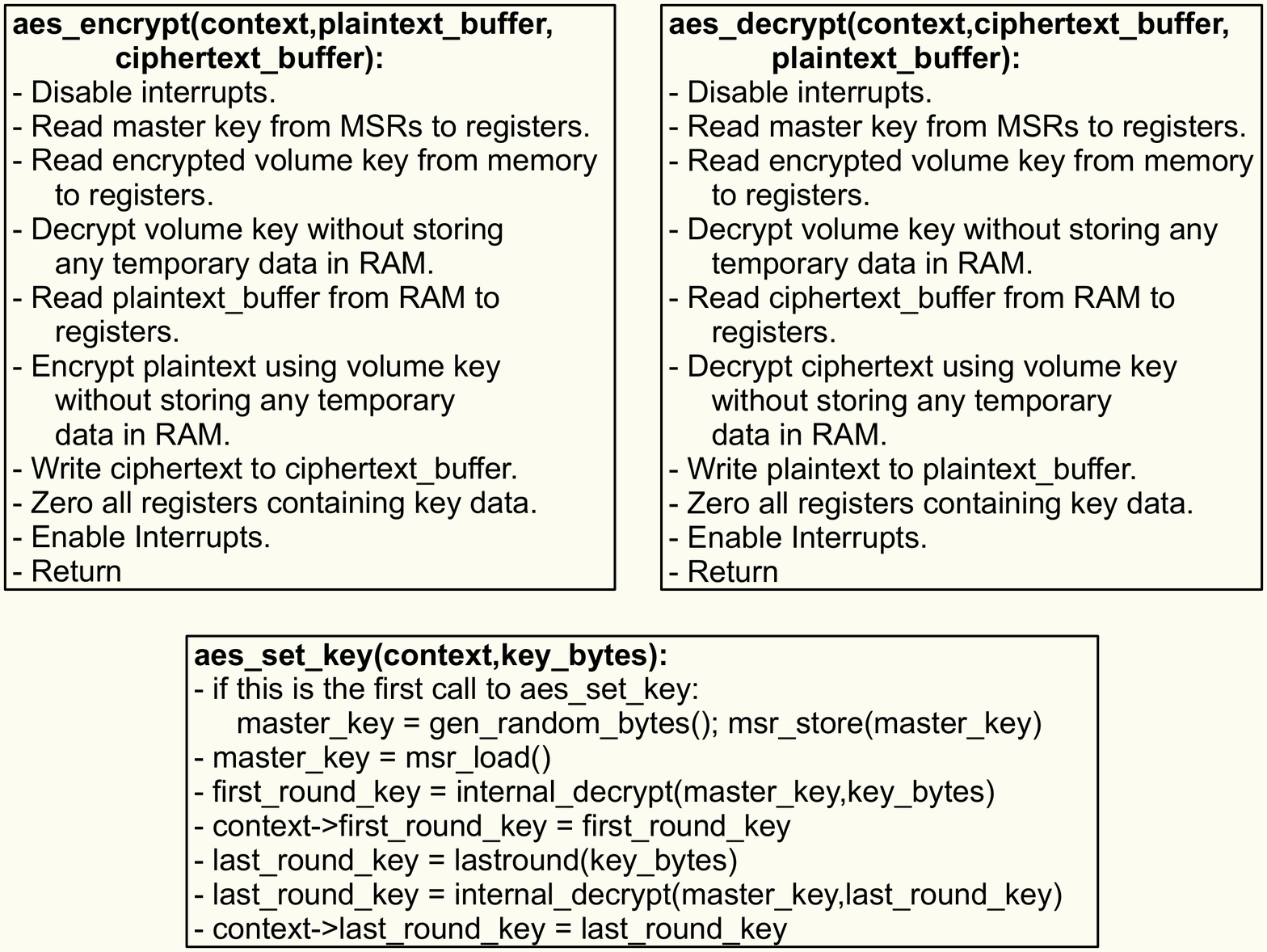}
\caption{Pseudocode Description of Loop-Amnesia}
\end{figure*}

\subsection{Constraints}

To validate our design, we built a cold-boot immune 128-bit AES
implementation as a drop-in replacement for the 128-bit AES
implementation already present in the loop-AES disk encryption
package.  In order to satisfy our primary design criterion of
cold-boot immunity, we must take care in our implementations of
\texttt{aes\_encrypt} and \texttt{aes\_decrypt} that no key data is
ever stored to RAM.  This places a number of constraints on our
implementation.

First, in order to ensure no register containing key data could ever
be spilled to RAM, we needed a degree of control over the register
allocation process not available to the programmer in any high-level
language, including C\@.  For this reason, our implementation of
Loop-Amnesia uses x86-64 assembly language exclusively.

Second, though most AES implementations, in order to improve
performance, precompute the AES key schedule and cache it to RAM, our
repurposed MSR space is far too limited to store even one full AES key
schedule.  We instead compute the key schedule on-the-fly during
encryption and decryption as discussed in \S\ref{outline}.

Finally, as MSRs are per-CPU (or per-core), the need to copy our
master key to all CPUs that may run the Loop-Amnesia subroutines
presents a logistical problem.  Our prototype implementation currently
handles this problem by compiling the Linux kernel in single-CPU mode,
forcing all software to execute on only one CPU or CPU core.  While
the prototype implementation of loop-AES therefore currently limits a
machine to a single core, there is nothing in the design of
Loop-Amnesia requiring this limitation.  In a production
implementation of Loop-Amnesia, we would suggest storing the master
key to RAM after its generation, forcing all CPUs to read it and store
it to their MSRs, and subsequently scrubbing the key from RAM.

\subsubsection*{The TPM Alternative}

Many of these design constraints could be lifted if hardware support
were available.  However, the Trusted Protection
Modules\cite{tpmfoolery} present on so many computers today do not
provide useful hardware support for our goal.  While it might at first
appear that we could secure the key inside of such a cryptographic
coprocessor and use it to perform all encryption and decryption of the
disk, the current TPM standard only supports the public-key RSA
algorithm, which is inappropriate for disk encryption.

However, even though TPMs are not useful for performing the actual
disk encryption, they could be used as an alternative method of
encrypting the disk volume keys: instead of using an AES key hidden in
an MSR on the main processor for the master key, we could use a public
RSA key generated by the TPM.  When we wanted to perform disk
encryption or decryption, we could ask the TPM to use the
corresponding private RSA key to decrypt the values we stored in RAM,
reading the decrypted disk volume key directly from the TPM to
registers over the serial bus.

Unfortunately, this is an inferior alternative to our approach from
both security and performance standpoints.  From a security
standpoint, the disk volume keys would frequently be transferred
unencrypted over a bus from the TPM to the system CPU.  An adversary
able to tap this bus would be able to obtain the disk volume keys.
From a performance standpoint, the master key would be decrypted by a
relatively slower algorithm on a relatively slower processor, and we
would in addition incur the latency of two transmissions over the
TPM-CPU bus for every volume key decryption.\footnote{Performance
  problems due to bus latency and TPM processor speed would plague
  even a hypothetical TPM implementation supporting AES or another
  symmetric encryption algorithm.}  For these reasons, we
chose not to utilize a TPM for our implementation.

\subsection{Implementation Outline}
\label{outline}

\begin{figure*}[t]
\includegraphics[scale=0.60]{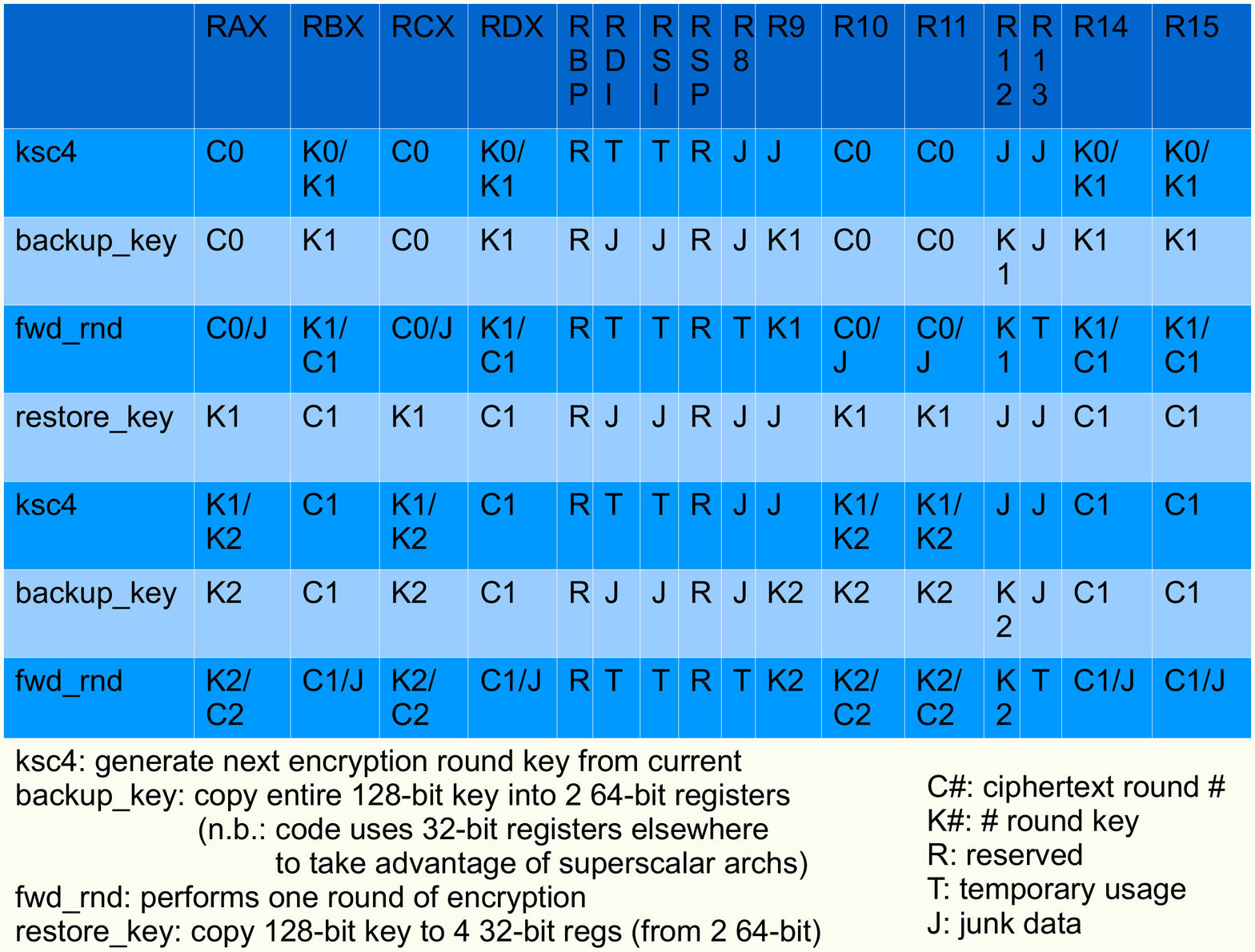}
\caption{Register Usage of Loop-Amnesia (2 rounds of 10 shown)}
\label{fig:reguse}
\end{figure*}



The \texttt{aes\_encrypt} and \texttt{aes\_decrypt} functions take an
AES context structure, a buffer containing the plaintext or
ciphertext, and a buffer to which the encrypted ciphertext or
decrypted plaintext must be stored.  Each of these functions must use
the master key to decrypt the volume key stored in the AES context
structure, use this decrypted key to encrypt the plaintext buffer or
decrypt the ciphertext buffer, and must finally write the fully
encrypted ciphertext or fully decrypted plaintext to the output
buffer.  Programming these cryptographic routines in assembly
language, on an architecture with 16 registers, and under the
constraint that RAM not be used for working storage proved,
predictably, to be a significant engineering challenge.

\texttt{aes\_encrypt} and \texttt{aes\_decrypt} work similarly as the
encryption and decryption operations are nearly symmetric.  There are
16 registers available for use on x86-64.  Of these, \texttt{RSP} is
the stack pointer and must always point to the stack, so it is not
available for our use.  We use \texttt{RBP} to point to the encryption
or decryption function, depending on which operation we wish to
perform.  The 16 bytes of partially encrypted plaintext or partially
decrypted ciphertext are moved from \texttt{EAX}, \texttt{ECX},
\texttt{R10D}, and \texttt{R11D} to \texttt{EBX}, \texttt{EDX},
\texttt{R14D}, and \texttt{R15D} during the performance of a single
round of encryption or decryption.  The routine performing a single
round of encryption or decryption uses \texttt{R8}, \texttt{R13},
\texttt{RDI}, and \texttt{RSI} as temporary registers.  The round key
is stored in \texttt{R9} and \texttt{R12} while each round is
performed.  See Figure \ref{fig:reguse} for an illustration of
Loop-Amnesia's register usage.

Thus, every general-purpose integer register available in the x86-64
instruction set is in use during the encryption and decryption
subroutines.  Since the 32-bit x86 architecture has only 8 integer
registers available, adapting this technique to 32-bit x86 would
likely require the use of the MMX or SSE registers.  Adopting the
technique to a RISC architecture with an abundance of general-purpose
registers, however, would be straightforward.


\texttt{aes\_set\_key} is the routine to initialize an AES context
structure with a given key.  Our implementation generates the master
key, if necessary, and initializes the AES context structure in RAM
with the first and last round keys, first encrypting each with the
master key.

\section{Verification of Cold-Boot Immunity}

\subsection{Justification}

A system will be immune to a cold-boot attack if, when the system is
running normally (i.e., not including directly after the input of a
key to the system), no key data is ever stored to RAM.  From the
perspective of the x86-64 assembler programmer, key data could
only be stored to RAM due to one of the following occurrences:
\begin{enumerate}
\item An explicit store, including a stack push instruction.
\item A taken interrupt causing registers with key data to be stored
  to the interrupt stack.
\end{enumerate}

A review of the code in \texttt{aes\_encrypt} and
\texttt{aes\_decrypt} easily shows that no register containing part of
any master key, volume key, or round key is ever stored to RAM.
Moreover, interrupts are disabled before the master key is read out of
the MSRs and only enabled after registers containing key data have all
been zeroed, so it is theoretically impossible for Loop-Amnesia to be
vulnerable to the cold-boot attack given its
structure.\footnote{Non-maskable interrupts, or NMIs, cannot be
  disabled by software, and it is therefore theoretically possible for
  key data to leak to RAM if NMIs must be considered.  We further
  discuss the problem of non-maskable interrupts in \S\ref{nmi}.}

While perhaps not strictly necessary for immunity to the cold-boot
attack, it is also not desirable that partially encrypted ciphertext
(such as after one round of encryption) be stored to RAM as an
attacker may be able to use cryptanalysis against such a degenerate
version of AES to recover the volume key.  Loop-Amnesia only stores
fully encrypted ciphertext or fully decrypted plaintext to RAM,
thwarting such an attack.

\subsection{Correctness Testing}

We performed correctness testing on an AMD Athlon64 X2 Dual Core
Processor 3800+\footnote{using only one core, for the reasons
  mentioned in \S\ref{implicate}}.  For convenience, we used the Linux
/dev/mem device to inspect the physical RAM of this machine, rather
than actually replicating the cold-boot attack ourselves.  Using this
methodology, we were able to extract the secret key from loop-AES.
When using Loop-Amnesia, we found neither the master key nor volume
key present in RAM.  We did, however, find data equivalent to the
volume key encrypted with the master key present in RAM, as we
expected.

\section{Performance}

\subsection{Benchmarking}

\begin{figure*}[t]
\includegraphics[scale=0.6]{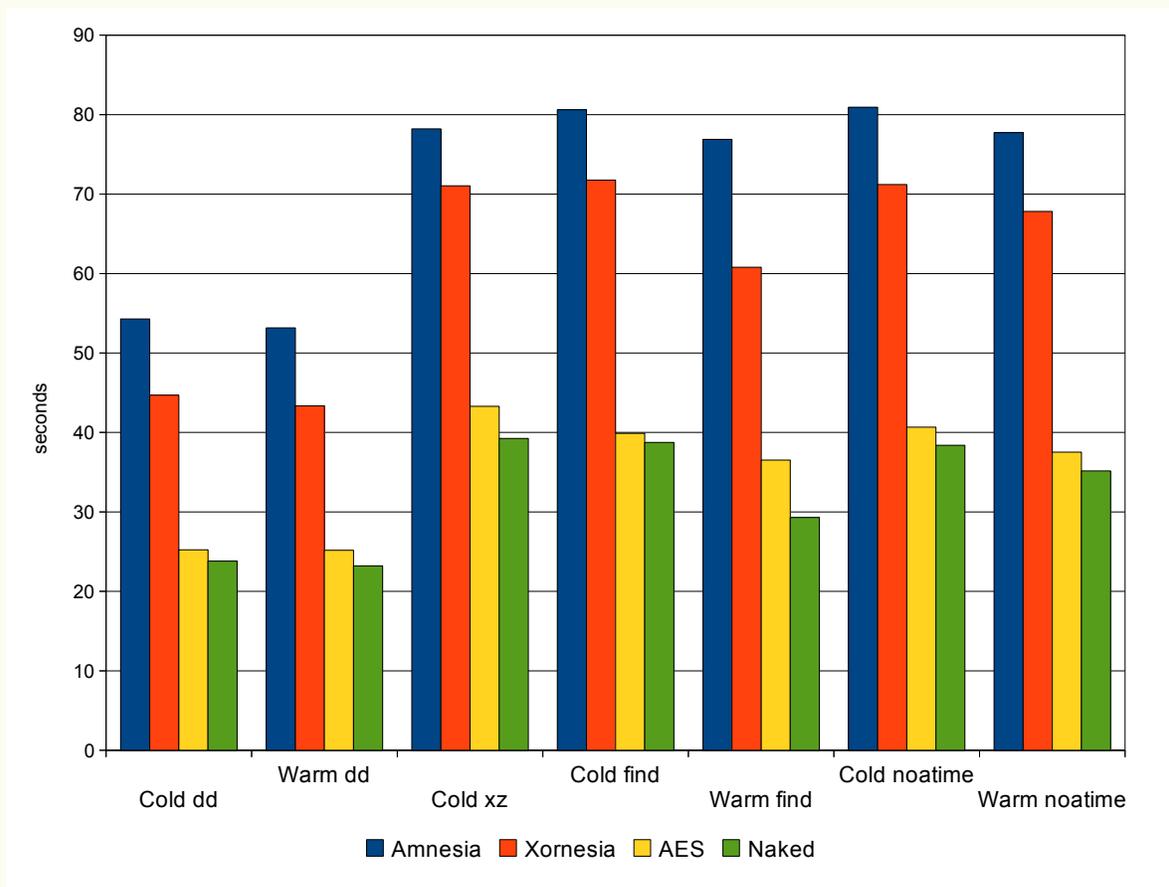}
\caption{Loop-Amnesia Performance} 
\label{fig:performance}
\end{figure*}

We compare Loop-Amnesia against three other disk encryption methods.
Our results are shown in Figure \ref{fig:performance}.  ``Xornesia''
refers to a modified version of Loop-Amnesia which encrypts the disk
volume keys in RAM by XORing them with the master key instead of
performing full AES.  Xornesia continues to use full AES when using
the disk volume keys to do encryption and decryption of user data.  We
use Xornesia to isolate the overhead caused by repeated calculation of
the key schedule, which is still present in Xornesia, from the
overhead caused by the need to repeatedly decrypt the disk volume
keys, which is not.  ``AES'' refers to the loop-AES 128-bit AES
implementation, with which we are fully compatible.  We use this to
measure the overhead of our Loop-Amnesia implementation relative to
state-of-the-art disk encryption software using an optimized
implementation of the same algorithm.  ``Naked'' refers to a simple
loopback mount with no encryption whatsoever.  We use this as our
baseline in order to eliminate from consideration the overhead of a
loopback device.

The benchmarks are small, disk-intensive shell operations.  dd writes
a 900MB file consisting entirely of zeroes to disk.  xz untars the
Linux kernel from an xz-format archive.  The ``find'' benchmark
searches the Linux kernel source tree for instances of a particular
word. ``noatime'' is the same as ``find'' but done on a filesystem
mounted with an option to disable the recording of the time of last
access.  ``Cold'' benchmarks are done with the disk cache cleared;
``warm'' benchmarks are done after the disk cache has been primed by
performing the same benchmark immediately before the test.  We do not
report numbers for warm xz as the CPU component of decompression made
this test a poor measure of disk performance.  We formatted the
encrypted loopback device with the ext2 filesystem for all tests and
used a single-core laptop with an Intel Celeron 540 at 1.8GHz with 1GB
of RAM for benchmarking.  The disk, a Hitachi HTS54258 (5400 RPM),
experimentally performs reads at 725MB/s from the disk cache (on CPU)
and at 44MB/s from the disk buffer (on disk microcontroller).  Our
results show that, on average, Loop-Amnesia introduces a slowdown of
approximately 2.04x relative to Loop-AES and 2.23x relative to an
unencrypted disk.

\begin{figure}[t]
\includegraphics[scale=0.35]{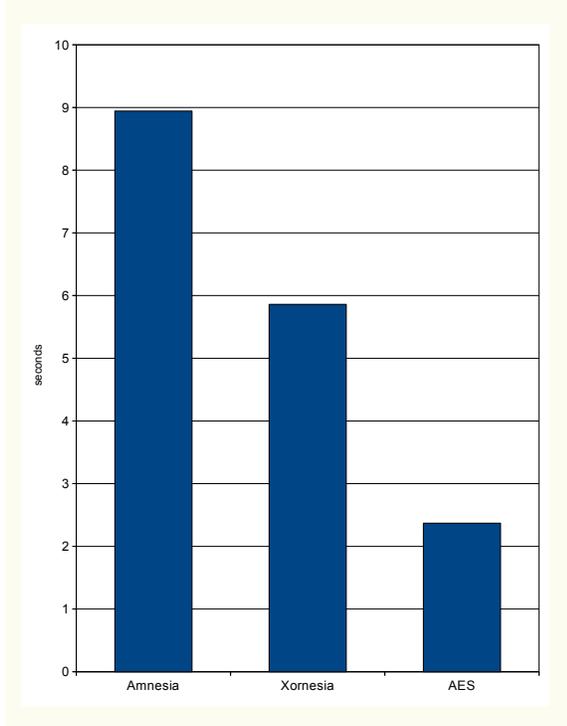}
\caption{Amnesia, Xornesia, and AES CPU time}
\label{fig:harness}
\end{figure}

We also ran a simple unit test pitting Loop-Amnesia, Xornesia, and
Loop-AES against each other, graphed in Figure \ref{fig:harness}.
Since this is a CPU test, not a test of performance in practice, this
provides a measure of the theoretical worst potential overhead
Loop-Amnesia could cause, which would occur if disk accesses were free
and performance of an encrypted filesystem was therefore bound
entirely by CPU speed.  The times given are for 10 million encryption
and decryption operations.  The theoretical worst-case slowdown of
Loop-Amnesia relative to Loop-AES was found to be 3.77x.

\subsection{Analysis}

While we would have preferred Loop-Amnesia to have less of a
performance impact, we believe that this overhead is acceptable given
the unique benefit we provide.  It is also worth noting that, while we
designed these benchmarks to stress the disk subsystem, disk access
speed does not play a major role in overall performance for many
computing applications.  The author has been using Loop-Amnesia for
several months on both the laptop used for conducting the benchmarks
and on another machine and has not noticed an appreciable decline in
performance on either machine for interactive desktop
use.\footnote{The machines did not previously use any form of disk
  encryption.}

Our overhead comes from two sources.  First, we must perform two
cryptographic operations for each single cryptographic operation we
are called on to perform by the loop-AES framework.  Specifically, we
must decrypt the device key with the master key, then use this
decrypted key to perform the cryptographic operation originally
requested (either encryption or decryption of a 16-byte block of
data).  Xornesia stores device keys XORed with the secret key rather
than performing AES to encrypt the device keys, therefore cutting out
the overhead of two cryptographic operations for every single act of
encryption or decryption.  Our second source of overhead is the
necessity of generating round keys on-the-fly; loop-AES pregenerates
these and keeps them within the AES context structure.

Though Xornesia has significantly lower overhead, we do not recommend
the use of Xornesia instead of our original algorithm as doing so
would weaken our security guarantee.  An adversary able to choose the
device key for an encrypted loopback device on the system would be
able to derive the master key by performing the cold-boot attack and
examining the encrypted device key.\footnote{It may also be possible
  to find the secret key by performing cryptanalysis on the first and
  last round keys in RAM, but we could negate this vulnerability by
  storing only the last round key in RAM and computing the first round
  key from the last whenever encryption is required.  This would still
  be faster than performing full AES.}  From this, the attacker could
discover the keys for encrypted loopback devices he did not configure.
We felt that our method of defeating the cold boot attack should
thwart even an attacker with user-level access to the machine.

\section{Limitations}

In this section, we discuss some limitations and potential
vulnerabilities of both our approach and of the current implementation
of Loop-Amnesia.

\subsection{Architecture Dependence}
Our approach is inherently architecture-dependent and limited to
encryption systems with a kernel-mode implementation.  An
assembly-language implementation must be completed for every
combination of CPU architecture and encryption algorithm needing
support.

However, we nevertheless feel our approach is applicable to a wide
variety of use cases.  Encryption algorithms are small, self-contained
pieces of code which only need be written once.  Our implementation
already supports a secure and widely used algorithm for the most
common desktop and server CPU architecture.  We expect that vendors
will have the resources to adapt their existing encryption algorithm
implementations -- which, as in the case of loop-AES, may have already
been implemented in assembly langauge for performance purposes -- to
use the Loop-Amnesia method for countering cold-boot attacks if there
is even moderate institutional demand.

\subsection{Functionality Limitations}

As the CPU registers, including the MSRs, are cleared when a computer
is suspended to RAM, we cannot support suspension to RAM.  It would be
possible for an implementation of our technique to copy the master key
to RAM before allowing the computer to suspend, but this would be
ill-advised: in such an implementation, the contents of the master key
would be at risk of discovery by a cold-boot attack if the attacker
gained access to the suspended computer.

\subsection{Potential Effectiveness Issues}

\paragraph{Espionage}
An attacker able to install a keystroke logger or otherwise tamper
with the victim computer may be able to deduce the key through
espionage.  While we do not protect against a keystroke logger, the
use of two-factor authentication, supported by loop-AES and
Loop-Amnesia, could reduce its effectiveness, and a trusted
path\cite{tpmfoolery} execution framework could be used to prevent an
attacker from tampering with unencrypted binaries used to mount the
encrypted disk.

\paragraph{Key Information in Userspace}
According to the developer of loop-AES, the userspace portions of the
cryptographic system of which Loop-Amnesia is a part will overwrite
userspace key material with zeroes after transmitting it to the
kernel\cite{ruusumail}.  However, since key material is transmitted
through a UNIX pipe, it may still be available in the buffer unless
the pipe is zeroed by the kernel after use; this is currently not
done.

\paragraph{Cached Data}
Large amounts of decrypted data may be cached to RAM by the operating
system, and our approach does not protect this data against a cold
boot attack.  However, it is possible for a user to manually clear the
Linux disk cache by writing to a special file\cite{dropcache}.
Periodically writing to this file from userspace, therefore, could
mitigate the effectiveness of this attack at the expense of
performance if the Linux kernel clears pages when they are freed
(instead of when they are allocated).  We have not checked whether the
Linux kernel does in fact clear freed pages, but it would be simple to
modify the operating system to do so.

\paragraph{JTAG}
Many processors implement a standardized debugging infrastructure
called the Joint Test Action Group, or JTAG.  By sending signals to a
CPU over JTAG, a hardware developer is able to test the CPU's
functioning.  JTAG is commonly used in verifying that a particular CPU
is not defective before releasing it for purchase.  Because it is
possible to use JTAG to dump the internal registers of a CPU, an
attacker able to access the JTAG debug port may be able to read the
Loop-Amnesia master key from the CPU's MSRs.  Fortunately, it is rare
for the JTAG debug port to be wired out for x86 processors\cite{celf}.
In the rare case that a JTAG port is available on an x86 machine, we
would recommend that a user concerned about this remove or destroy the
JTAG port and/or blow the JTAG security fuse.  Either of these actions
would disable an attacker's ability to access JTAG\cite{jtagfuse}.

\paragraph{Non-Maskable Interrupts}
\label{nmi}
We take care to disable interrupts before reading the master key into
general-purpose registers and to reenable them only after the key has
once again been erased from all general-purpose registers.  However,
some interrupts, called non-maskable interrupts (NMIs), cannot be
disabled.  These interrupts are usually caused only by hardware
faults.  Since the general-purpose registers are stored to RAM when an
interrupt is taken, an attacker able to introduce a hardware fault
during the brief time periods when key material is in the
general-purpose registers would be able to read the master key.  We
consider such an attack unlikely to prove practical, primarily due to
its complexity and dependence on extreme luck in timing.  However, if
this attack does prove to be a concern, modifying the operating
system's interrupt handler to scrub the general-purpose registers from
RAM after receiving a non-maskable interrupt would be sufficient to
protect against it.  This would have no deleterious side effects as
the hardware will have faulted, so the CPU will never resume normal
execution.\footnote{Our prototype implementation does not modify the
  OS interrupt handlers.}

\section{Related Work}


\subsection{Lest We Remember: Cold-Boot Attacks on Encryption Keys}

Halderman et al.\ discussed some forms of mitigation in
\cite{princetonattack}, including deleting keys from memory when an
encrypted drive is unmounted\footnote{this is already done in loop-AES
  according to \cite{ruusumail}}, obfuscation techniques, and hardware
modifications such as intrusion-detection sensors and epoxy-encased
RAM.  Halderman et al.\ admit that they do not present a full solution
applicable to general-purpose hardware.

While special-purpose hardware modifications may be effective, such
hardware adds cost and may not be available to many users of disk
encryption; a solution for commodity hardware is required.  As the
cold-boot attacker is given a copy of all RAM, including the program
text used to perform encryption and decryption, we doubt that
obfuscation would prove effective.

\subsection{AESSE}

A paper at Eurosec 2010 \cite{aesse} discussed a potential solution to
the cold boot attack, in which a single encryption key was stored in
the MMX registers of the CPU and MMX register access was disabled for
user-level code.  Encryption can then be performed by using MMX or SSE
instructions in kernel mode to perform AES encryption or decryption.
The method proposed causes an algorithmic performance slowdown of
approximately 6x.  In addition to having worse performance
characteristics than Loop-Amnesia, AESSE also does not support
multiple disk encryption keys, since only one encryption key schedule
may be stored inside the MMX registers.  Disabling access to the MMX
registers also causes compatibility problems with userland software
that requires MMX and performance slowdowns for userland software that
would make use of MMX if available but cannot because of AESSE.

\subsection{Braving the Cold Black Hat Talk}

A talk\cite{blackhat} at Black Hat in 2008 discussed various methods
of mitigating the effects of the cold boot attack.  Most of these
mitigation strategies are discussed elsewhere; however, one
contribution of this talk is a suggestion that motherboard temperature
sensors be used to detect attempts to cool RAM and take protective
measures, such as scrubbing the keys.

This talk also proposed a potential solution to the cold boot attack.
The researchers suggested that the key could be stored in RAM only as
the product of the hash of a large block of bits.  The hope is that at
least one of these bits will flip during the performance of the cold
boot attack, preventing its success.  This strategy, if implemented,
would likely suffer from severe performance problems as a large hash
would need to be calculated every time an encryption key needed to be
accessed.  The talk also discussed ``caching'' the encryption key
inside the MMX registers, but it was unclear from the talk how such a
caching system would operate.

\subsection{Frozen Cache}

J\"{u}rgen Pabel has posted a website\cite{frozencache}, dormant since
early 2009, detailing his plans to provide a software-based solution
to the cold-boot attack.  His approach is to memory-map the L1 cache
of the CPU and use this space to store the AES key schedule.  Because
this approach would prevent the CPU cache from serving its normal
role, every memory access on the machine would result in a cache miss.
Disabling the CPU cache in this manner results in a slowdown of
perhaps 200x\cite{cachegrind} felt by all software, not just software
accessing files on the encrypted disk.  Because our solution avoids
the negative system performance side effects of Pabel's design, we
believe it to be more practical.

\subsection{Linux-Crypto Mailing List Brainstorming}

Shortly after Halderman et al.\ published their attack, a mailing
list discussion on Linux-Crypto discussed possible mitigation
strategies.  The general approach of keeping key information in CPU
registers was brought up\cite{armchair}, but the ideas given were too
vague to suggest how this might specifically be accomplished and do
not appear to have been pursued further.

\subsection{Leakage-Resistant Algorithms}

There has been considerable work \cite{Akavia09simultaneoushardcore}\cite{DziembowskiP08}\cite{Naor2009}\cite{leakpractice}
in designing cryptosystems resilient to partial key leakage due to
side channel attacks.  Most of this work has focused on the design of
new ciphers with properties mitigating the impact of partial key
leakage.

Unfortunately, we do not believe that protecting against partial key
leakage is a sufficient defense against the cold boot attack.
According to Halderman et al., it is possible to perform the cold-boot
attack in such a way that over 99.9\% of memory remains uncorrupted an
entire minute after power is cut.  Any countermeasure to the cold-boot
attack must account for its potential to fully leak any encryption
keys stored to RAM.

\subsection{TCG Platform Reset Attack Mitigation Specification}

The Trusted Computing Group has published a
standard\cite{tcgmitigation} which purports to mitigate the
vulnerability of compliant systems to the cold-boot attack.  This
specification states that a compliant BIOS must zero out all RAM
before giving control to the operating system.  While this prevents
the attack from being performed using only the victim's
computer\footnote{A BIOS password would also necessitate the use of a
  separate machine.}, the attacker can still easily perform the attack
by moving the RAM from the victim's machine to a machine under his own
control, then booting using a BIOS not following the TCG
specification.  Thus, the TCG specification cannot be considered a
sufficient countermeasure.

\subsection{Forenscope: A Framework for Live Forensics}

The RAM of a computer may contain sensitive material other than the
encryption keys to the hard disk.  The Forenscope rootkit
\cite{forenscope} takes advantage of the cold-boot attack to gain
access to active network sessions as well; the session keys' presence
in memory could allow an attacker to masquerade as the victim to any
website, SSH server, or other remote system to which the user was
connected at the time of the attack.

Loop-Amnesia will protect against a Forenscope-using attacker's
gaining access to the encrypted disk: the attack tool uses the exact
same strategy as Halderman et al.\ to attempt recovery of the key.
Unfortunately, SSH and SSL session keys will likely remain in RAM, so
an attacker with Forenscope could still conceivably keep the victim's
network connections alive, sniff the session keys, and masquerade as
the victim to connected machines.  See \S\ref{thefuture} for a
discussion on how Loop-Amnesia may be extended to assist in preventing
Forenscope attacks.

\section{Future Work}
\label{thefuture}

A ripe area for future research is the applicability of our approach
to algorithms outside of the AES cipher family.  Some algorithms, such
as Blowfish\cite{blowfish}, use key-dependent S-boxes; proving whether
these S-boxes can be safely stored to RAM would require careful
analysis.  We believe that our approach should work well for all
algorithms without key-dependent S-boxes and with key schedules that
are computationally inexpensive to compute, but its effectiveness
outside this class of ciphers remains to be analyzed.

The ability of Loop-Amnesia to assist in neutralizing Forenscope's
other attack capabilities also merits examination.  For instance, an
operating system attempting to harden itself against Forenscope could
use Loop-Amnesia to encrypt various pieces of data inside the kernel
TCP stack.  As the master key will have been erased by the reboot
preceding Forenscope's installation, Forenscope will have no way of
recovering the network connections.  By the time the attacker has had
time to download and analyze the SSH/SSL session keys from RAM, any
active TCP sessions will likely have expired.

Finally, our work exposes a limitation in current system programming
languages: the inability to insist to a compiler that particular
values never be spilled to RAM.  While we recognize that our needs are
uncommon and do not by themselves merit the redesign of system
programming languages, we speculate that programming language
designers may one day wish to allow users more control over the
register allocation process for performance reasons.  We would
encourage the designers of such languages or language extensions to
include functionality allowing the user to express the needs we faced
when implementing Loop-Amnesia.  User control over the register
allocation process may provide useful benefits for both security and
performance.

\section{Conclusion}

In this paper, we present the first practical solution to the
cold-boot attack applicable to general-purpose hardware.  For a
performance cost likely to be very moderate under most workloads, our
solution provides protection for general-purpose hardware against a
significant practical attack affecting all previous state-of-the-art
disk encryption systems.  We present a design strategy applicable to
all operating system-based disk encryption systems and a usable
open-source
implementation which
validates our design.  After the publication of this paper, we intend
to work with the Linux kernel community to integrate our approach, and
possibly code, into the standard Linux kernel distribution.

\section{Acknowledgements}

We thank Andrew Lenharth of the University of Texas at Austin for
his invaluable inspiration and advice in the early stages of this
work.  We also thank Jari Ruusu for providing loop-AES to the free and
open source software community: being able to use such well-designed
software as the base for our implementation significantly aided us in
evaluating the concepts behind Loop-Amnesia.

\bibliographystyle{plain}
\bibliography{amnesia}

\begin{thebibliography}{10}

\bibitem{cachegrind}
Cachegrind: a cache-miss profiler.
\newblock \url{http://wwwcdf.pd.infn.it/valgrind/cg_main.html}.

\bibitem{truecrypt}
Truecrypt: Free open-source on-the-fly encryption.
\newblock \url{http://www.truecrypt.org/}.

\bibitem{bbcscary}
Hard drive secrets sold cheaply.
\newblock \url{http://news.bbc.co.uk/2/hi/technology/3788395.stm}, June 2004.

\bibitem{dropcache}
drop\_caches.
\newblock \url{http://www.linuxinsight.com/proc_sys_vm_drop_caches.html}, May
  2006.

\bibitem{dumbdoc}
Privacy at risk after burglary at doctor's office.
\newblock
  \url{http://www.cbc.ca/health/story/2011/01/21/nb-privacy-warning.html},
  January 2011.

\bibitem{Akavia09simultaneoushardcore}
Adi Akavia, Shafi Goldwasser, and Vinod Vaikuntanathan.
\newblock Simultaneous hardcore bits and cryptography against memory attacks.
\newblock In {\em Theory of Cryptography Conference}, pages 474--495, 2009.

\bibitem{celf}
Mike Anderson.
\newblock Using a {JTAG} in linux driver debugging.
\newblock In {\em CE Embedded Linux Conference}, 2008.
\newblock \url{http://elinux.org/images/4/4e/CELF_JTAG_Anderson.ppt}.

\bibitem{cryptoeprint:2009:317}
Alex Biryukov and Dmitry Khovratovich.
\newblock Related-key cryptanalysis of the full {AES}-192 and {AES}-256.
\newblock Cryptology ePrint Archive, Report 2009/317, 2009.
\newblock \url{http://eprint.iacr.org/}.

\bibitem{bureaucracy}
Bob Brown.
\newblock How to roll out full disk encryption on your pcs and laptops.
\newblock \url{http://www.networkworld.com/news/2010/081610-encryption.html},
  August 2010.

\bibitem{forenscope}
E.~Chan, S.~Venkataraman, F.~David, A.~Chaugule, and R.~Campbell.
\newblock Forenscope: A framework for live forensics.
\newblock In {\em Annual Computer Security Applications Conference}, November
  2010.

\bibitem{inteldoc}
Intel Corporation.
\newblock {IA-32} architectural {MSR}s.
\newblock {\em Intel 64 and IA-32 Architectures Software Developer's Manual},
  3B:681--722, January 2011.
\newblock \url{http://www.intel.com/Assets/PDF/manual/253669.pdf}.

\bibitem{windows}
Microsoft Corporation.
\newblock Bitlocker drive encryption technical overview.
\newblock {\em Microsoft Technet}, 2010.
\newblock
  \url{http://technet.microsoft.com/en-us/library/cc732774\%28WS.10\%29.aspx}.

\bibitem{sva}
John Criswell, Andrew Lenharth, Dinakar Dhurjati, and Vikram Adve.
\newblock Secure virtual architecture: a safe execution environment for
  commodity operating systems.
\newblock In {\em Proceedings of Twenty-First ACM SIGOPS Symposium on Operating
  Systems Principles}, SOSP '07, pages 351--366, New York, NY, USA, 2007. ACM.

\bibitem{boucher}
John Curran.
\newblock Encrypted laptop poses 5th amendment dilemma.
\newblock {\em USA Today}, February 2008.
\newblock
  \url{http://www.usatoday.com/tech/news/techpolicy/2008-02-07-encrypted-lapto%
p-child-porn_N.htm}.

\bibitem{rijndael}
Joan Daemen and Vincent Rijmen.
\newblock {\em The Design of Rijndael}.
\newblock Springer-Verlag New York, Inc., Secaucus, NJ, USA, 2002.

\bibitem{amddoc}
Advanced~Micro Devices.
\newblock {MSR}s of the {AMD}64 architecture.
\newblock {\em AMD64 Architecture Programmer's Manual}, 2:469--472, June 2010.
\newblock \url{http://support.amd.com/us/Processor_TechDocs/24593.pdf}.

\bibitem{DziembowskiP08}
Stefan Dziembowski and Krzysztof Pietrzak.
\newblock Leakage-resilient cryptography.
\newblock In {\em FOCS}, pages 293--302, 2008.

\bibitem{linuxrngcounter}
Jake Edge.
\newblock Holes in the linux random number generator?
\newblock {\em Linux Weekly News}, 2006.
\newblock \url{http://lwn.net/Articles/184925/}.

\bibitem{wackenhut}
David~W. Foley.
\newblock \url{http://doj.nh.gov/consumer/pdf/wackenhut.pdf}, December 2010.

\bibitem{cryptoeprint:2009:531}
Henri Gilbert and Thomas Peyrin.
\newblock Super-sbox cryptanalysis: Improved attacks for aes-like permutations.
\newblock Cryptology ePrint Archive, Report 2009/531, 2009.
\newblock \url{http://eprint.iacr.org/}.

\bibitem{tcgmitigation}
Trusted~Computing Group.
\newblock {TCG} platform reset attack mitigation specification.
\newblock
  \url{http://www.trustedcomputinggroup.org/resources/pc_client_work_group_pla%
tform_reset_attack_mitigation_specification_version_10/}, 2008.

\bibitem{linuxrngvuln}
Zvi Gutterman, Tzachy Reinman, and Benny Pinkas.
\newblock Analysis of the linux random number generator.
\newblock In {\em IEEE Symposium on Security and Privacy}, 2006.

\bibitem{princetonattack}
J.~Alex Halderman, Seth~D. Schoen, Nadia Heninger, William Clarkson, William
  Paul, Joseph~A. Calandrino, Ariel~J. Feldman, Jacob Appelbaum, and Edward~W.
  Felten.
\newblock Lest we remember: Cold boot attacks on encryption keys.
\newblock In Paul~C. van Oorschot, editor, {\em USENIX Security Symposium},
  pages 45--60. USENIX Association, 2008.

\bibitem{jtagfuse}
Zack~Albus Markus~Koesler, Franz~Graf.
\newblock Programming a flash-based msp430 using a {JTAG} interface.
\newblock \url{http://www.softbaugh.com/downloads/slaa149.pdf}, December 2002.

\bibitem{blackhat}
Patrick McGregor, Tim Hollebeek, Alex Volynkin, and Matthew White.
\newblock Braving the cold: New methods for preventing cold boot attacks on
  encryption keys, 2008.

\bibitem{aesse}
Tilo M\"{u}ller, Andreas Dewald, and Felix~C. Freiling.
\newblock Aesse: a cold-boot resistant implementation of aes.
\newblock In {\em Proceedings of the Third European Workshop on System
  Security}, EUROSEC '10, pages 42--47, New York, NY, USA, 2010. ACM.

\bibitem{Naor2009}
Moni Naor and Gil Segev.
\newblock Public-key cryptosystems resilient to key leakage.
\newblock In {\em Proceedings of the 29th Annual International Cryptology
  Conference on Advances in Cryptology}, pages 18--35, Berlin, Heidelberg,
  2009. Springer-Verlag.

\bibitem{frozencache}
J\"{u}rgen Pabel.
\newblock \url{http://frozencache.blogspot.com}, 2009.

\bibitem{solaris}
OpenSolaris Project.
\newblock {ZFS} on-disk encryption support.
\newblock \url{http://hub.opensolaris.org/bin/view/Project+zfs-crypto/WebHome}.

\bibitem{loopaes}
Jari Ruusu.
\newblock \url{http://loop-aes.sourceforge.net/}.

\bibitem{ruusumail}
Jari Ruusu.
\newblock \url{http://mail.nl.linux.org/linux-crypto/2008-06/msg00002.html},
  June 2008.

\bibitem{linux}
Christophe Sauot.
\newblock dm-crypt: A device-mapper crypto target.
\newblock \url{http://www.saout.de/misc/dm-crypt/}.

\bibitem{blowfish}
Bruce Schneier.
\newblock Description of a new variable-length key, 64-bit block cipher
  (blowfish).
\newblock In {\em Fast Software Encryption, Cambridge Security Workshop}, pages
  191--204, London, UK, 1994. Springer-Verlag.

\bibitem{freescaledoc}
Freescale Semiconductor.
\newblock Performance monitor counter registers.
\newblock {\em MPC750 RISC Processor Family User's Manual}, pages 378--382,
  December 2001.
\newblock
  \url{http://www.freescale.com/files/32bit/doc/ref_manual/MPC750UM.pdf}.

\bibitem{secvisor}
Arvind Seshadri, Mark Luk, Ning Qu, and Adrian Perrig.
\newblock Sec{V}isor: a tiny hypervisor to provide lifetime kernel code
  integrity for commodity {OS}es.
\newblock {\em SIGOPS Oper. Syst. Rev.}, 41:335--350, October 2007.

\bibitem{leakpractice}
Francois-Xavier Standaert, Olivier Pereira, Yu~Yu, Jean-Jacques Quisquater,
  Moti Yung, and Elisabeth Oswald.
\newblock Leakage resilient cryptography in practice.
\newblock In David Basin, Ueli Maurer, Ahmad-Reza Sadeghi, and David Naccache,
  editors, {\em Towards Hardware-Intrinsic Security}, Information Security and
  Cryptography, pages 99--134. Springer Berlin Heidelberg, 2010.

\bibitem{tpmfoolery}
Allan Tomlinson.
\newblock Introduction to the {TPM}.
\newblock
  \url{http://courses.cs.vt.edu/cs5204/fall10-kafura-BB/Papers/TPM/Intro-TPM-2%
.pdf}.

\bibitem{hypersafe}
Zhi Wang and Xuxian Jiang.
\newblock Hypersafe: A lightweight approach to provide lifetime hypervisor
  control-flow integrity.
\newblock In {\em IEEE Symposium on Security and Privacy}, pages 380--395,
  2010.

\bibitem{armchair}
Richard Zidlicky.
\newblock \url{http://www.spinics.net/lists/crypto/msg04668.html}, 2008.

\end{thebibliography}

\end{document}